\documentclass[two-column, nofootinbib]{revtex4-1}

\usepackage{amsmath}
\usepackage{tabularx}
\usepackage{graphicx}
\usepackage[usenames,dvipsnames,svgnames]{xcolor}
\usepackage{hyperref}

\hypersetup{dvips,dvipdfm,colorlinks=true,urlcolor=magenta,filecolor=magenta,linktoc=page,citecolor=red,linkcolor=blue,bookmarks=true}

\begin{document}

\title{Cosmological thermodynamics with Hawking temperature on the apparent horizon and Unruh temperature of the fluid: Some interesting consequences}

\author{Subhajit Saha\footnote{Electronic Address: \texttt{\color{blue} subhajit1729@gmail.com}}}
\affiliation{Department of Mathematics, Jadavpur University, Kolkata 700032, West Bengal, India \\ and \\
Department of Physical Sciences, \\ Indian Institute of Science Education and Research Kolkata, Mohanpur 741246, West Bengal, India.}


\begin{abstract}

\begin{center}
(Dated: The $14^{\text{th}}$ July, $2016$)
\end{center}

Thermodynamics on the cosmological apparent horizon of a flat Friedmann-Lemaitre-Robertson-Walker metric has been investigated with Bekenstein entropy and Hawking temperature on the horizon, and Unruh temperature for the fluid inside the horizon. This temperature is experienced by a radial comoving observer infinitesimally close to the horizon due to the pressure exerted by the fluid bounded by the horizon. An expression for the entropy of the fluid has been obtained which is found to be proportional to the volume of the thermodynamic system  which implies that the Unruh temperature of the fluid is inconsistent with the holographic principle. Further, we have been able to find an expression for the effective entropy of the system. Finally, assuming a barotropic equation of state $p=w\rho$ ($w$ constant) for the fluid, it has been shown that the generalized second law holds good for a non-phantom $w$, while thermodynamic equilibrium is never possible for such a scenario.

\end{abstract}

\maketitle



Research in the theory of black holes (BHs) in General Relativity during the past 40 years has revealed a very deep and fundamental relationship among gravitation, thermodynamics, and quantum theory. The cornerstone of this relationship is BH thermodynamics, where it appears that at the purely classical level, BHs obey certain laws which bear a remarkable mathematical resemblance to the ordinary laws of thermodynamics (for details, one may see Ref. \cite{Wald1}). In the semiclassical description of BH physics, it has been found that a BH behaves as a black body and emits thermal radiation. The temperature (known as Hawking temperature \cite{Hawking1}) and the entropy (known as Bekenstein entropy \cite{Bekenstein1}) are proportional to the surface gravity at the horizon and the area of the horizon respectively. The laws of BH thermodynamics have been generalized in the context of cosmology and considering the Universe filled with a cosmic fluid and bounded by some horizon to be an isolated thermodynamic system, there have been a lot of work \cite{Wang1, Izquierdo1, Chakraborty1, Saha1, Saha2, Chakraborty2} investigating the validity of the thermodynamic laws, particularly the first law (FL), the generalized second law (GSL), and thermodynamic equilibrium. 

The apparent horizon of the Universe always exists and the thermodynamical properties related to the apparent horizon have been studied by several authors including in a quasi-de Sitter geometry of inflationary universe \cite{Frolov1} and a late time accelerating Q-space \cite{Bousso1}. A common assumption in these kind of studies is that the temperature of the apparent horizon and that of the fluid bounded by it are equal, i.e., the Hawking temperature. However, it seems that this assumption may not be valid\footnote{Another possible reason for this assumption to be flawed is that since we are considering an isolated thermodynamic system, so an exchange of energy between the fluid and the horizon is possible only if they have different temperatures.} for all kinds of cosmic fluids. This will become clear if we carefully look into the variation of fluid temperature with the cosmic expansion. Assuming no non-gravitational interaction, the temperature falls \cite{Padmanabhan1, Weinberg1, Padmanabhan2, Rahvar1} with expansion [$a(t)$ is the scale factor of the Universe] as $1/a$ for relativistic fluids such as photons (radiation) and as $1/a^2$ for non-relativistic fluids like dust and cold dark matter (CDM). Even if there is any interaction, these expressions get only slightly modified because the said interaction, if at all, is expected to be of very small strength. Further, considering Hawking temperature for the fluid, an exact expression for the fluid entropy has never been obtained. 

In this Letter, we shall consider a flat Friedmann-Lemaitre-Robertson-Walker (FLRW) universe and investigate the validity of the thermodynamic laws, particularly the GSL and thermodynamic equilibrium, on the apparent horizon assuming that the temperature of the fluid is the Unruh temperature. Using Gibb's law, we shall attempt to determine an expression for the fluid entropy and hence the effective entropy of the thermodynamic system. Finally, assuming a barotropic equation of state (EoS) $p=w\rho$ ($w$ constant) for the fluid inside the apparent horizon, we shall show that the GSL holds good for a non-phantom $w$, but the system never attains thermodynamic equilibrium.

Let us consider a flat FLRW universe with the line element
\begin{equation}
ds^2=-dt^2+a^2(t)\left[dr^2+r^2(d\theta ^2 +\text{sin}^2\theta d\phi ^2)\right].
\end{equation}
The Einstein field equations are
\begin{equation} \label{efe}
H^2=\frac{8\pi G}{3}\rho~~~~~~~~\text{and}~~~~~~~~\dot{H}=-4\pi G(\rho +p),
\end{equation}
from which one can derive the conservation equation as
\begin{equation} \label{ce}
\dot{\rho}+3H(\rho +p)=0.
\end{equation}
In Eqs. (\ref{efe}) and (\ref{ce}), $H=\frac{\dot{a}(t)}{a(t)}$ is the Hubble parameter, $\rho$ is the total energy density of the cosmic fluid, and $p$ is the pressure of the fluid.

If $R=ar$ is the area radius, then for a flat universe, the location of the dynamical apparent horizon is given by
\begin{equation}
R_A=\frac{1}{H}
\end{equation}
The velocity of the apparent horizon can be calculated as
\begin{equation} \label{va}
\dot{R}_{A}=-\frac{\dot{H}}{H^2}=1+q,
\end{equation}
where $q$ is the deceleration parameter of the Universe. A positive $q$ indicates deceleration and a negative $q$ denotes acceleration.

The thermodynamic parameters, namely, entropy (Bekenstein) and temperature (Hawking) on the apparent horizon are given by
\begin{equation}
S_A=\left(\frac{c^3}{G\hbar}\right)\pi R_{A}^{2}
\end{equation}
and
\begin{equation}
T_{A}^{(h)}=\left(\frac{\hbar c}{\kappa _B}\right)\frac{1}{2\pi R_A}
\end{equation}
respectively. In the above equations, $\hbar$ is the reduced Planck constant and $\kappa _B$ is the Boltzmann's constant. For simplicity in our calculations, without any loss of generality, we shall assume the quantities $8\pi$, $G$, $c$, $\hbar$, and $\kappa _B$ to be unity. Note that the Hawking temperature $T_{A}^{(h)}$ is suitable for relativistic fluids, so it is relevant in a radiation dominated universe only (i.e., when $p=\frac{1}{3}\rho$). It is unsuitable for a dust (or CDM) dominated universe, ergo there is a need to search for an expression for the temperature of such non-relativistic fluids. 

Now according to thermodynamics, the equilibrium configuration of an isolated macroscopic physical system should
be the maximum entropy state, consistent with the constraints imposed on the system. Thus if $S$ is the entropy of the system, the following conditions should hold --- (a) $dS \geq 0$ [i.e., the entropy function cannot decrease (the second law of thermodynamics)], and (b) $d^2 S < 0$ [i.e., the entropy function attains a maximum (thermodynamic equilibrium)]. In our context, the Universe bounded by the apparent horizon and filled with some cosmic fluid forms an isolated macroscopic physical system for which the above inequalities can be generalized as
\begin{equation}
\text{(i)}~d(S_A+S_f) \geq 0 ~~~~~~~~\text{and}~~~~~~~~ \text{(ii)}~d^2 (S_A+S_f) < 0
\end{equation}
respectively, where $S_f$ is the entropy of the fluid bounded by the horizon. The inequality (i) is sometimes called the GSL.


The differential of the entropy $S_A$ on the apparent horizon can be obtained as
\begin{equation} \label{dsa}
dS_A=\frac{1}{4}R_A\dot{R}_A dt
\end{equation}
It is generally assumed in these situations that the fluid inside the horizon bears the same temperature as that on the horizon. In that case, from Gibb's relation \cite{Izquierdo1}
\begin{equation} \label{grel}
T_{A}^{(h)}dS_f=dE_f+pdV_h,
\end{equation}
where $E_f=\rho V_h$ is the total energy of the fluid inside and $V_h=\frac{4}{3}\pi R_{h}^{3}$ is the volume of the fluid bounded by the horizon, we obtain the expression for the differential of the fluid entropy as
\begin{equation} \label{dsf}
dS_f=\frac{1}{8}R_{A}^{4}(\rho +p)(\dot{R}_A-1)dt.
\end{equation} 
Taking the sum of Eqs. (\ref{dsa}) and (\ref{dsf}), and using Eq. (\ref{va}), the differential of the total entropy (or the effective entropy) can be written as
\begin{equation} \label{dtot}
d(S_A+S_f)=\frac{1}{4}R_A \dot{R}_A \lbrace 1+R_A (\dot{R}_A-1) \rbrace dt. 
\end{equation}
It can be observed that the right hand side of Eq. (\ref{dtot}) cannot be expressed as a perfect differential. So, it is not possible to obtain an expression for the effective entropy $(S_A+S_f)$. Such an expression can be obtained if somehow the factor $\dot{R}_A-1$ (which equals $q$) in Eq. (\ref{dsf}) is eliminated. In the following, we shall show that the Unruh temperature for the fluid bounded by the apparent horizon leads to interesting consequences as far as the thermodynamics of the system is concerned.


The Unruh effect, or sometimes the Fulling-Davies-Unruh effect, is the prediction that an accelerating observer will observe black­body radiation where an inertial observer would observe none. This effect was first described by S.A. Fulling \cite{Fulling1}, P.C.W. Davies \cite{Davies1}, and W.G. Unruh \cite{Unruh1}. The expression for the Unruh temperature is given by
\begin{equation}
T_u=\left(\frac{\hbar}{\kappa _B c}\right)\frac{a_r}{2\pi},
\end{equation}
with 
\begin{center}
$a_r=-\frac{\ddot{a}}{a}R$
\end{center}
as the acceleration experienced by a radial comoving observer at $r$, namely, at the place of the horizon (or infinitesimally close to the horizon). The negative sign arises because we consider that the acceleration is caused by the matter (fluid) in the spatial region enclosed by the horizon. Note that the proper acceleration vanishes for a comoving observer. This acceleration may therefore be considered as a "pseudo acceleration" experienced by a comoving observer caused by the pressure exerted by the fluid bounded by the horizon. In this regard, it should also be mentioned that Unruh temperature has been considered in the literature in the context of entropic cosmology \cite{Cai1}.

In relativistic units, the Unruh temperature becomes
\begin{equation} \label{unruht}
T_u=-\frac{\ddot{a}R}{2\pi a}.
\end{equation}
Note that this temperature (expressed in Kelvin scale) is positive when the Universe decelerates, while it is negative during cosmic acceleration. Negative temperatures are interesting and somewhat unusual, but not impossible or paradoxical. They are related to the concept of population inversion \cite{Reif1, Singh1} in statistical physics. The population inversion is obtained by what is called optical pumping, which is a process of imparting energy to the working substance of a laser to transfer the atoms to excited states. Systems with a negative temperature will decrease in entropy as one adds energy to the system \cite{Atkins1}. Most familiar systems cannot achieve negative temperatures because adding energy always increases their entropy. The possibility of decreasing in entropy with increasing energy requires the system to "saturate" in entropy\footnote{This implies that systems with negative absolute temperatures deviates from thermodynamic equilibrium.}, with the number of high energy states being small. These kinds of systems, bounded by a maximum amount of energy, are generally forbidden classically. Thus, negative temperature is a strictly quantum phenomenon. 
Further, one can easily verify that for non-relativistic fluids such as dust or CDM, $T_u \propto 1/a^2$.

It is quite interesting to note that the ratio of the Unruh temperature to the Hawking temperature for the apparent horizon equals the deceleration parameter $q$ \cite{Mitra1}, i.e.,
\begin{equation}
\frac{T_{A}^{(u)}}{T_{A}^{(h)}}=q.
\end{equation}
Thus replacing $T_{A}^{(h)}$ by $T_{A}^{(u)}(=qT_{A}^{(h)})$ in Eq. (\ref{grel}), the expression for the fluid entropy can be evaluated as
\begin{eqnarray} \label{dsfu}
dS_f &=& \frac{1}{8}R_{A}^{4}(\rho +p)dt \nonumber \\
&=& \frac{1}{4}R_{A}^{2}\dot{R}_Adt,
\end{eqnarray}
which allows us to express the right hand side as a perfect differential, i.e.,
\begin{equation}
dS_f=d\left(\frac{1}{12}R_{A}^{3}\right).
\end{equation}
So, we have in fact obtained an expression for the fluid entropy given by
\begin{equation}
S_f=\frac{1}{12}R_{A}^{3},
\end{equation}
which shows that the entropy of the fluid is proportional to the volume of the system. Note that this expression is independent of the nature of the cosmic fluid.

The differential of the total entropy can now be written as
\begin{equation} \label{deff}
d(S_A+S_f)=\frac{1}{4}R_A\dot{R}_A(1+R_A)dt,
\end{equation}
from which we can get the effective entropy as
\begin{equation}
S_{eff}=S_A+S_f=\frac{1}{24}(3R_{A}^{2}+2R_{A}^{3}).
\end{equation}


Now, if the cosmic fluid is assumed to be a perfect fluid having a barotropic EoS $p=w\rho$ ($w$ constant), then the deceleration parameter $q$ has the expression
\begin{equation}
q=\frac{1}{2}(1+3w).
\end{equation}
Then, the differential of the effective entropy becomes [from Eq. (\ref{deff})]
\begin{eqnarray}
dS_{eff} &=& \frac{1}{4}R_A(1+R_A)(1+q)dt \nonumber \\
&=& \frac{3}{8}(1+w)R_A(1+R_A)dt,
\end{eqnarray}
which shows that the GSL holds good for a non-phantom EoS parameter $w$. The second order differential of $S_{eff}$ can be evaluated as
\begin{equation}
\frac{d^2}{dt^2}(S_{eff})=\frac{9}{16}(1+w)^2(1+2R_A),
\end{equation}
thus leading us to conclude that the thermodynamic equilibrium for such a system can never be attained, as expected from our discussion in the paragraph after Eq. (\ref{unruht}).

Summarizing this Letter, a thermodynamic study of the apparent horizon has been done with Hawking temperature and Bekenstein entropy on the horizon but Unruh temperature of the fluid contained within the apparent horizon. This temperature is thought to be experienced by a radial comoving observer infinitesimally close to the horizon due to the pressure exerted by the fluid bounded by the horizon. We have considered a flat FLRW universe for the purpose and as far as the apparent horizon is concerned, the Unruh temperature equals the Hawking temperature multiplied by the deceleration parameter of the Universe. Using Gibb's law, we have succeded in obtaining an expression for the entropy of the fluid inside the apparent horizon and hence an effective entropy of the thermodynamic system. Further, the fluid entropy is proportional to the volume of the thermodynamic system and the effective entropy is independent of the nature of the chosen cosmic fluid. The expression for the fluid entropy shows that our consideration of Unruh temperature for the fluid bounded by the horizon is inconsistent with the holographic principle. Finally, assuming a barotropic EoS of the fluid having a constant EoS parameter, we have evaluated the first and second order differentials of the effective entropy. We have been able to conclude that although the GSL holds good for a non-phantom EoS, but the thermodynamic equilibrium for such a system is forbidden.

Finally, it is worthwhile to note that negative temperature states have been demonstrated in localized systems with finite, discrete spectra. Further, Braun et al. \cite{Braun1} had prepared a negative temperature state for motional degrees of freedom. It could be shown that our atomic system is stable, even though the atoms strongly attract each other --- that means they want to collapse but cannot due to being at negative temperature. Negative temperatures therefore imply negative pressures and open up new parameter regimes for cold atoms, enabling fundamentally new many-body states. Our Universe as a whole is also not collapsing under the attractive force of gravity, rather it is undergoing an accelerated expansion. Dark energy (DE), which is believed to possess a huge negative pressure, has been introduced to describe this effect. 
It seems that negative absolute temperatures might play an important role in the dynamics of our Universe and whether further investigations reveal a deep connection between the nature of DE and the temperature of the cosmic fluid remains to be seen.


\begin{acknowledgments}

The author is grateful to Diego Pavon, Stefano Viaggiu, and Alexis Helou for their insightful comments and criticisms on an earlier version of the manuscript. The author also thanks the anonymous referees for their comments which have helped to enhance the quality of the manuscript. This work was supported by SERB, Govt. of India under National Post-doctoral Fellowship Scheme [File No. PDF/2015/000906].

\end{acknowledgments}




\end{document}